\documentstyle [psfig,aps]{revtex}

\begin{document}
\draft
\title{On the Nature of the Generic Big Bang}
\author{Beverly K. Berger}
\address{Department of Physics, Oakland University, Rochester, MI  48309, USA}

\maketitle
\bigskip
\begin{abstract}
Spatially homogeneous but possibly anisotropic cosmologies have two main types of 
singularities: (1) asymptotically velocity term dominated (AVTD)---(reversing the 
time 
direction) the universe evolves to the singularity with fixed anisotropic collapse 
rates ; (2) Mixmaster---the anisotropic collapse rates change in a deterministicaly chaotic
way. Much less is known about spatially inhomogeneous universes. It has been  claimed that
a generic universe would evolve toward the singularity as a different Mixmaster universe at
each spatial point. I shall discuss how to predict whether a cosmology has an AVTD or
Mixmaster singularity and whether or not our numerical simulations agree with these
predictions.
\end{abstract}
\section{Introduction}
EinsteinÕs theory of general relativity has passed all experimental tests to date and 
is thus, almost certainly, the classical theory of the gravitational field. Yet the
universality and nonlinearity of the theory have been proven (Hawking(1970)) to imply that
regular but generic initial data will (for reasonable classical matter) develop a
singularity either to the future or the past. In the simplest examples, this singularity is
hidden from us. The Friedmann-Robertson-Walker  homogeneous, isotropic cosmology has a
spacelike big bang hidden from our view by the cosmic microwave background surface of last
scattering. Spherically symmetric black holes have a spacelike internal singularity hidden
within an event horizon. Spatially homogeneous but  anisotropic (different collapse rates
in different directions) cosmologies have two types of singularities. An asymptotically
velocity term dominated (AVTD) singularity means that the cosmology eventually behaves as a
Kasner spacetime (fixed anisotropic expansion rates  characterized by a single parameter
$u$) (Isenberg and Moncrief (1990)).  The other type of singularity is called Mixmaster
(Belinskii et al (1971), Misner(1969)). Here the spatial scalar curvature creates a
potential such that a bounce off the potential corresponds to a change of Kasner solution.
The parameter  $u$ changes in a deterministic way which is generally believed to be
chaotic. The nature of the singularities in generic spacetimes is largely unknown. Recent
analytic and  numerical results demonstrate that some spatially inhomogeneous cosmologies
approach a different Kasner singularity at each spatial point and thus have AVTD
singularities. Belinskii, Khalatnikov, and Lifshitz (1971) (BKL) long ago conjectured that
the generic singularity is locally Mixmaster-like. While this conjecture remains
controversial, it provides a paradigm which can be tested.

 In this talk, I shall discuss the progress toward understanding the nature of the 
generic 
cosmological singularity (Berger, Garfinkle, and Moncrief (1997)).  In particular, I shall
emphasize the  application of a method due to Grubi\u{s}i\'{c} and Moncrief
(1993) (GM) to predict whether one should expect the singularity to be
AVTD or Mixmaster-like and then discuss numerical studies to support or contradict such
predictions. This work was done in collaboration with David  Garfinkle (Oakland
University), Vincent Moncrief (Yale University), James Isenberg (University of Oregon) and
graduate students Eugene Strasser (Oakland University),  Boro Grubi\u{s}i\'{c} (Yale
University), and Marsha Weaver (University of Oregon). 

\section{ Mixmaster dynamics}
To understand the BKL conjecture, let us first discuss a vacuum Bianchi IX cosmology. 
According to the Bianchi classification of 3-dimensional homogeneous spaces, the spatial 
line element has the form
\begin{equation}
dl^2=A_{ij}\sigma ^i\sigma ^j 
\end{equation}
where the $A_{ij}$
 are constants and the $\sigma ^i$Ôs (generalizations of $dx^i$) satisfy
\begin{equation}
d\sigma ^i=C^i_{jk}\,\sigma ^j\wedge \sigma ^k. 
\end{equation}
For (vacuum) Bianchi IX, $C^i_{jk}=\varepsilon ^i_{jk}$, the antisymmetric symbol. The 
spatial dependence of the $\sigma ^i$Ôs when written in a coordinate basis yield a
non-vanishing spatial scalar curvature, $^3R$.  To obtain a cosmological spacetime, we
shall let the $A_{ij}$ become functions of time only. For convenience, we shall require
$A_{ij}$ to be diagonal. It is then useful to represent these anisotropic scale factors
$e^\alpha $,
$e^\zeta $, and $e^\gamma $ in terms of the logarithmic volume $\Omega$ and orthogonal 
anisotropic shears $\beta _\pm $ where
\begin{eqnarray}
\alpha &=&\Omega -2\beta _+, \nonumber \\
  \zeta &=& \Omega +\beta _++\sqrt 3\beta _-, \nonumber \\
  \gamma &=& \Omega +\beta _+-\sqrt 3\beta _-.  
\end{eqnarray}
EinsteinÕs equations can be obtained by variation of the Hamiltonian constraint
\begin{equation}
H=-p_\Omega ^2+p_+^2+p_-^2+V(\beta _+,\beta _-,\Omega ) 
\end{equation}
where $p_\Omega $, $p_\pm $ are conjugate to $\Omega $, $\beta _\pm $ and
\begin{equation}
V(\beta _+,\beta _-,\Omega ) =e^{4\Omega -8\beta _+}+e^{4\Omega +4\beta
_++4
\sqrt 3\beta _-}+e^{4\Omega +4\beta _+-4\sqrt 3\beta _-} 
 -2e^{4\Omega +4\beta _+}-2e^{4\Omega -2\beta _+-2\sqrt 3\beta _-}-2e^{4\Omega -2
\beta _++2\sqrt 3\beta _-}. 
\end{equation}
The singularity occurs as $\Omega \to -\infty $. Note that Eq.\ (4) splits naturally into 
a kinetic piece	
\begin{equation}
H_K=-p_\Omega ^2+p_+^2+p_-^2 
\end{equation}
and a potential piece. EinsteinÕs equations also require 
\begin{equation}
H=0. 
\end{equation}
We shall now explore the method of consistent potentials (MCP) as used by GM. 
\begin{figure}[bth]
\begin{center}
\makebox[4in]{\psfig{file=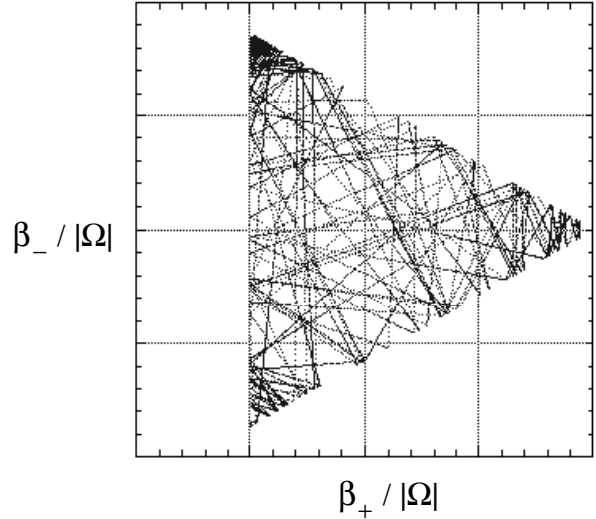,width=3.5in}}
\caption{A typical Mixmaster trajectory as projected into the anisotropy plane. The
rescaled variables (Moser et al (1971)) cause the bounces to occur at fixed locations. More
than 262 bounces are shown (see Berger, Garfinkle, and Strasser (1997)).}
\end{center}
\end{figure}
The case $V\equiv 0$ corresponds in fact to the Kasner solution (Bianchi Type I with 
$C^i_{jk}=0$). Thus Eq.\ (4) may be replaced by $H_K$ whose variation yields in the 
limit $\Omega \to -\infty $,
\begin{equation}
\beta _\pm ={{p_\pm } \over {|p_\Omega |}}|\Omega | . 
\end{equation}
However, we require from Eq.\ (7) (now $H_K=0$)
\begin{equation}
p_\Omega ^2=p_+^2+p_-^2 
\end{equation}
and can replace Eq.\ (8) with 
\begin{equation}
\beta _+=\cos \theta |\Omega |\quad ,\quad \beta _-=\sin \theta |\Omega | .
\end{equation}
If the Mixmaster model were AVTD, substitution of Eq.\ (10) into Eq.\ (5)
would yield only  terms which decay exponentially as $\Omega \to -\infty $. 
 
Consider the first term on the right hand side (rhs) of Eq.\ (5). Upon substitution with 
Eq.\ (10), we find
\begin{equation}
e^{4\Omega -8\beta _+}\to e^{-4|\Omega |(1+2\cos \theta )}. 
\end{equation}
If $\cos \theta <-{\textstyle{1 \over 2}}$, this term will grow exponentially. A bounce 
off this exponential wall then occurs leading to a new value for $\cos \theta $. 
Conservation of momentum yields (Ryan (1971))
\begin{equation}
\cos \theta '=-{{4+5\cos \theta } \over {5+4\cos \theta }}
\end{equation}
which causes the first term on the rhs of Eq.\ (5) to exponentially decay. However, 
depending on the sign of $\sin \theta $, either the second or third term on the rhs of
Eq.\ (5) will begin to grow exponentially. This process repeats indefinitely. Numerical
studies  (Moser et al (1973), Rugh and Jones (1990), Berger, Garfinkle, and Strasser
(1997)) confirm this picture (see Fig.\ 1). This means that all terms which were neglected
in the MCP are in fact negligible.

\section{Gowdy universes on $T^3 \times R$}
If spatial dependence in one direction (e.g. on $0\le \theta \le 2\pi $) is allowed in 
the anisotropic scale factors of a Bianchi I homogeneous cosmology, the $T^3\times R$ Gowdy
(1971) solution is obtained (see also Berger (1974)). The metric
\begin{equation}
ds^2=e^{-\lambda / 2}e^{\tau / 2}(-e^{-2\tau }d\tau ^2+d\theta
^2)+e^{-\tau }[e^Pd\sigma ^2+2e^PQ\,d\sigma \kern 1pt d\delta +(e^PQ^2+e^{-P})d\delta 
^2]
\end{equation}
describes amplitudes $P(\theta ,\tau )$ and $Q(\theta ,\tau )$ for the $+$ and $\times$ 
polarizations of gravitational waves propagating in an inhomogeneous background spacetime
written in terms of $\lambda (\theta ,\tau )$.  This model has the extremely nice property
that the dynamical equations for the waves obtained by variation of the (non-zero)
Hamiltonian (Moncrief (1981))
\begin{equation}
H=\oint {d\theta \left( {{\textstyle{1 \over 2}}\pi
_P^2+{\textstyle{1 
\over 2}}e^{-2P}\pi _Q^2} \right)} 
+\oint {d\theta \,\left( {{\textstyle{1 \over 2}}e^{-2\tau }P,_\theta
^2+{\textstyle{1 
\over 2}}e^{2(P-\tau )}Q,_\theta ^2} \right)}=H_K+H_V 
\end{equation}
decouple from the constraints.  Here $\pi _P$ and $\pi _Q$ are the momenta conjugate to
$P$ and $Q$ respectively.  The constraints are first order equations for $\lambda$ which
can be solved trivially in terms of the $P$ and $Q$ found as solutions to the wave
equations obtained from Eq.\ (14). 
    
To use the MCP, we obtain the AVTD solution as $\tau \to \infty $ by variation of $H_K$ 
from Eq.\ (14). This yields 
\begin{equation}
P\to v(\theta )\tau \quad ,\quad Q\to Q_0(\theta ) .  
\end{equation}
Consider substitution in the separate terms of Eq.\ (14). The potential
\begin{equation}
V_1=\pi _Q^2e^{-2P} 
\end{equation}
in $H_K$ becomes
\begin{equation}
V_1=\pi _Q^2e^{-2v\tau } 
\end{equation}
which decays exponentially only for $v>0$. On the other hand,
\begin{equation}
V_2=Q,_\theta ^2e^{2(P-\tau )} 
\end{equation}
in $H_V$ becomes
\begin{equation}
V_2=Q,_\theta ^2e^{2(v-1)\tau } 
\end{equation}
which decays exponentially if $v<1$.  Thus we see that $0<v<1$ (at each spatial point) is 
consistent as $\tau \to \infty $ with an AVTD singularity.
\begin{figure}[bth]
\begin{center}
\makebox[4in]{\psfig{file=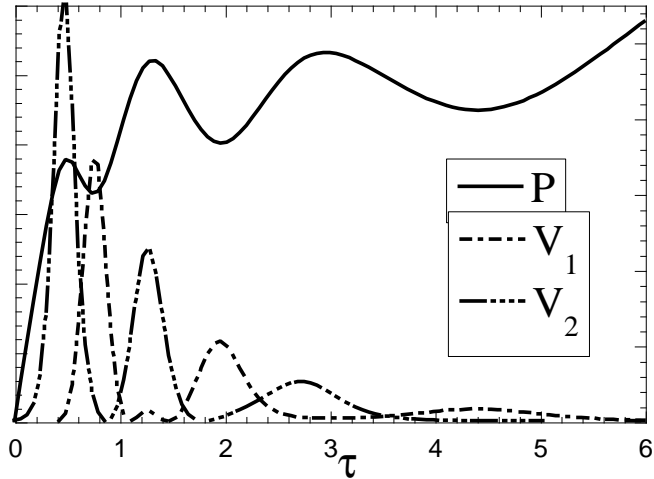,width=3.5in}}
\caption{$P$, $V_1$, and $V_2$ vs $\tau$ at a fixed value of $\theta$.  Note that the 
slope of $P$, $P,_\tau $, decreases after each interaction with $V_2$ while $P,_\tau $ goes
from negative to positive after each interaction with  $V_1$.  A continuation of this graph
in $\tau$ would show that $V_1$ and $V_2$ have permanently died off and that $P$ continues
to increase with fixed positive slope $P,_\tau <1$.}
\end{center}
\end{figure}
What happens if $P,_\tau <0$ or $P,_\tau >1$? Then either $V_1$ or $V_2$ will be large. A 
bounce off each potential will drive $P,_\tau $ into the range allowed by that potential.
Alternate bounces off both potentials will eventually drive $P,_\tau $ into the required
range. This process is shown at a typical spatial point in Fig.\ 2 which is taken from a
numerical simulation (Berger and Moncrief (1993), Berger and Garfinkle (1997)).
     
We note that exceptional behavior can occur at isolated spatial points where a coefficient 
of $V_1$ or $V_2$ (i.e. $\pi _Q$ or $Q,_\theta $) vanishes. Since near these exceptional
points $V_1$ or $V_2$ is small, the mechanism to drive $P,_\tau $ into the range $(0,1)$
takes a very long time to become effective. This means that $P,_\tau >1$ or $P,_\tau <0$
can persist leading to the development of spiky features in $P$ and $Q$.  These are
discussed in detail elsewhere (Berger and Garfinkle (1997)).
     
The MCP combined with numerical simulations provide strong evidence that the Gowdy 
singularity is AVTD except, perhaps, at isolated spatial points. 

\section{$U(1)$ symmetric cosmologies on $T^3 \times R$}
We have also considered a class of cosmologies with dependence on two spatial variables 
$u$, $v$ as well as time $\tau$.  The model may be described by five degrees of freedom
$\{z,x,\varphi ,\omega ,\Lambda \}$ with conjugate momenta $\{p_z,p_x,p_\varphi ,p_\omega
,p_\Lambda \}$.  The metric takes the form (Moncrief (1986), Berger and Moncrief (1993))
\begin{equation}
ds^2=e^{-2\varphi }\left\{ {-N^2e^{-4\tau }d\tau ^2+e^{-2\tau }e^\Lambda
e_{ab}dx^adx^b} 
\right\}+e^{2\varphi }\left( {dx^3+\beta _adx^a} \right)^2 
\end{equation}
where $\{x^a \}= \{u,v \}$, $N=e^\Lambda $, $x^3$ is the symmetry direction and
\begin{equation}
e_{ab}={\textstyle{1 \over 2}}\left(
{\matrix{{e^{2z}+e^{-2z}(1+x)^2}&{e^{2z}+e^{-2z} (x^2-1)}\cr
{e^{2z}+e^{-2z}(x^2-1)}&{e^{2z}+e^{-2z}(1-x)^2}\cr
}} \right) 
\end{equation}
is the conformal metric in the $u$-$v$ plane. The variable $\omega$ is obtained from a 
canonical transformation of the twists $\beta ^a$.  Einstein's equations are obtained from
the variation of 
\begin{eqnarray}
H&=&\oint {\oint {dudv}}\left( {{\textstyle{1 \over 8}}p_z^2+{\textstyle{1
\over  2}}e^{4z}p_x^2+{\textstyle{1 \over 8}}p_\varphi ^2+{\textstyle{1 \over
2}}e^{4\varphi }p_\omega ^2-{\textstyle{1 \over 2}}p_\Lambda ^2+2p_\Lambda } \right)
 +e^{-2\tau }\oint {\oint {dudv}}\left\{ {\left( {e^\Lambda e^{ab}}
\right),_{ab}-\left(  {e^\Lambda e^{ab}} \right),_a\Lambda ,_b} \right. \nonumber \\
  & &+e^\Lambda \left[ {\left( {e^{-2z}} \right),_ux,_v-\left( {e^{-2z}} \right),_vx,_u} 
\right] +\left. {2e^\Lambda e^{ab}\varphi ,_a\varphi ,_b+{\textstyle{1 \over 2}}e^\Lambda 
e^{-4\varphi }e^{ab}\omega ,_a\omega ,_b} \right\}=H_K+H_V. 
\end{eqnarray}
Unfortunately, the constraints are nontrivial. So far, we only have a restricted solution 
to them initially and have not yet implemented a program to enforce them during numerical
evolution. (Constraint preservation is guaranteed analytically but not in a differenced
approximation to EinsteinÕs equations.)
     
Despite the complicated appearance of Eq.\ (22), the MCP may still be used. Variation of 
$H_K$ in Eq.\ (22) yields as $\tau \to \infty $ the AVTD solution
\begin{eqnarray}
z\to -v_z(u,v)\tau \quad ,\quad x\to x_0(u,v)\quad ,\quad \varphi 
\to -v_\varphi (u,v)\tau \quad ,\quad \omega \to \omega _0(u,v)\quad , \nonumber \\
  \Lambda \to \Lambda _0(u,v)+[2-p_\Lambda ^0(u,v)]\tau \quad ,\quad p_z\to 
-4v_z(u,v)\quad ,\quad p_x\to p_x^0(u,v)\quad , \nonumber \\
  p_\varphi \to -4v_\varphi (u,v)\quad ,\quad p_\omega \to p_\omega ^0(u,v)\quad ,
\quad p_\Lambda \to p_\Lambda ^0(u,v)\quad .
\end{eqnarray}
In the AVTD limit, the Hamiltonian constraint becomes
\begin{equation}
\tilde H_K=2v_z^2+{\textstyle{1 \over 2}}e^{-4v_z}(p_x^0)^2+2v_\varphi
^2+{\textstyle{1 
\over 2}}e^{-4v_\varphi }(p_\omega ^0)^2-{\textstyle{1 \over 2}}(p_\Lambda ^0)^2=0 .
\end{equation}
The exponential terms in $\tilde H_K$ decay only for $v_z>0$ and $v_\varphi
>0$ as in the Gowdy spacetime and will act to drive $z,_\tau $ and $\varphi ,_\tau $ into
the allowed range. We require $p_\Lambda ^0>0$ to enforce collapse. Thus we need only
consider the restriction 
\begin{equation}
\tilde H_K=2v_z^2+2v_\varphi ^2-{\textstyle{1 \over 2}}(p_\Lambda ^0)^2=0
\end{equation}
as  $\tau \to \infty $.  Although $H_V$ is complicated, there are only two
types of  exponential dependence. Schematically, for $H_V=\oint {\oint {dudv}}\,V$, 
\begin{equation}
V\approx Ce^{\Lambda -2\tau -2z}+e^{\Lambda -2\tau -2z-4\varphi }(\nabla
\omega )^2 . 
\end{equation}
\begin{figure}[bth]
\begin{center}
\makebox[4in]{\psfig{file=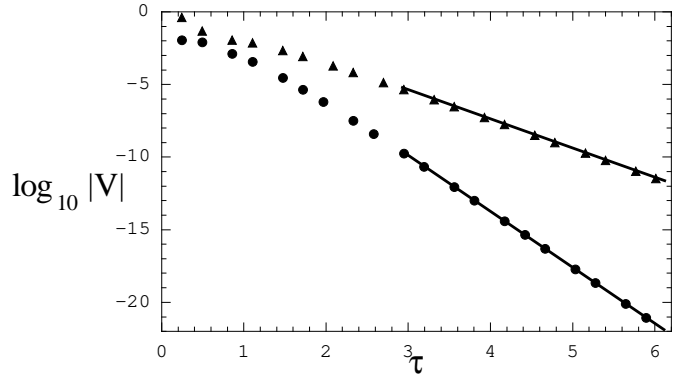,width=3.5in}}
\caption{The logarithm of the non-schematic $V$ (from Eq.\ (22)) vs $\tau$ for a polarized
$U(1)$ model. The solid lines are exponential best fits.}
\end{center}
\end{figure}
Only the term in Eq.\ (22) containing spatial derivatives of $\omega$ has the behavior
shown in the  second term on the rhs of Eq.\ (26). The others all behave as the first term.
Substitution of the AVTD solution Eq.\ (23) and the restriction given by Eq.\ (25) yield
(again schematically)
\begin{equation}
V\approx Ce^{-2v_\varphi \tau }+e^{2v_\varphi \tau }(\nabla \omega )^2 .
\end{equation}
The special case of polarized $U(1)$ models have $\omega =p_\omega =0$. 
This is the only  degree of freedom that may be permanently removed (both analytically and
numerically). In this case,
\begin{equation}
V \approx Ce^{-2v_\varphi \tau } . 
\end{equation}
Thus we expect polarized models to have AVTD singularities (Grubi\u{s}i\'{c} and Moncrief
(1994)).  The numerical simulations verify this prediction (Berger, Garfinkle, and
Moncrief (1997), Berger and Moncrief (1997)). Fig.\ 3 shows the exponential decay of $V$ at
two typical spatial points. Behaviors of
$\Lambda$,
$z$, $x$, and $\varphi$ are also consistent with an AVTD singularity.
     
For generic models, Eq.\ (27) predicts a growing term in $V$. This means that if a 
collapsing spacetime tries to become AVTD, a potential term will grow exponentially. A
bounce off this potential will (among other things) change the sign of $\varphi ,_\tau $ so
that an exponential term in $\tilde H_K$ will grow. The process should repeat indefinitely
since neither  $v_\varphi >0$ nor $v_\varphi <0$ is consistent with both terms
exponentially small.
\begin{figure}[bth]
\begin{center}
\makebox[4in]{\psfig{file=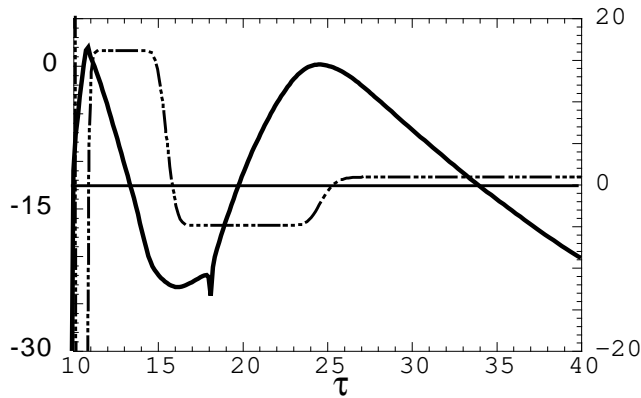,width=3.5in}}
\caption{The potential $V$ vs $\tau$  The base-10 logarithm of the (non-schematic---i.e.
from Eq.\ (22)) second term on the rhs  of Eq.\ (27) is plotted (solid line, left scale).
Note the exponentially growing and decaying portions of the graph.  Also plotted (broken
line) is the combination of momenta $(p_\Lambda + p_\varphi + p_z / 2)$ that yields the
negative of the quantity that multiplies
$\tau$ in the AVTD limit. The horizontal line indicates zero for this quantity (right
scale). }
\end{center}
\end{figure}

Results from a simulation (Fig.\ 4) appear to be consistent with this picture which itself 
is in accord with the BKL conjecture. However, there is a large {\it caveat}. The exponents
in $V$ in Eq.\ (26) are regulated by the kinetic part of the Hamiltonian constraint 
Eq.\ (25). If the value of
$\tilde H_K$ is incorrect due to numerical error, the exponent may have the wrong sign
yielding qualitatively wrong behavior. Thus these models will not provide information on
the validity of the BKL conjecture until enforcement of the constraint is implemented. In
polarized $U(1)$ models, the result is much stronger because the constraints remain small
(and converge to zero with increasing spatial resolution) for all $\tau $ (Berger and
Moncrief (1997))). 
     
A cleaner example of very similar behavior in one spatial dimension will be discussed 
elsewhere (Weaver et al (1997)).

\section{Conclusions}
We explore the behavior of cosmological spacetimes in their collapse to a singularity. 
The MCP can predict whether the singularity will be AVTD or Mixmaster-like. In the Gowdy
case, we learn how the nonlinearities in the wave equations force the parameters into the
range for consistent AVTD behavior. At isolated spatial points, these terms are absent so
non-generic behavior can occur. This leads to the growth of spiky features.
    
 In polarized $U(1)$ cosmologies, these spiky features do not arise. Numerical evolutions 
support the prediction that these models are AVTD. In generic $U(1)$ models, we expect a
Mixmaster-like singularity. Numerical difficulties prevent a definitive conclusion in this
case although there is some evidence to support this claim.
    
 We emphasize that, to the extent we believe the numerical results, the character of the 
singularity is completely local. The predictions of the MCP assume constancy in time of the
spatially dependent coefficients of the exponentially growing or decaying terms.
Simulations support such behavior sufficiently close to the singularity.
     
While the singularity occurs at (say) $\tau =\infty $, we have performed numerical 
simulations only to some $\tau _F<\infty $. Nonetheless, we believe that we can use these
simulations to support conjectures about the nature of singularities in these models. Since
the numerical simulations support predictions based on the MCP, we argue that we understand
the effect of all the nonlinear terms on the evolution toward the singularity. It seems
reasonable to suppose that for $\tau _F<\tau <\infty $ nothing qualitatively different will
happen. This is almost certainly true for AVTD spacetimes. An infinite sequence of
Mixmaster bounces at different spatial points could yield eventually a very complicated
spatial structure (e.g. Kirillov and Kochnev (1987)). It is likely, nevertheless, that the
exponential potentials will still dominate over increasing spatial derivatives as the
controlling factor in the evolution.

\section*{Acknowledgments}
This work was supported in part by National Science Foundation Grant PHY9507313 to Oakland 
University. I would like to thank the Institute for Geophysics and Planetary Physics of
Lawrence Livermore National Laboratory for hospitality.


\begin{references}
\bibitem{bkl}
Belinskii, V. A., Lifshitz, E. M. and Khalatnikov, I. M. (1971), Oscillatory Approach to 
the Singular Point in Relativistic Cosmology, {\it Sov. Phys. Usp}., {\bf 13}:  74--765.

\bibitem{berger1}
Berger, B. K.  (1974), Quantum Graviton Creation in a Model Universe, {\it Ann. Phys. 
(N.Y.)}, {\bf 83}:  458.

\bibitem{berger2}
Berger, B. K. and Garfinkle, D. (1997), Phenomenology of the Gowdy Model on $T^3\times R$, 
gr-qc/9710102.

\bibitem{berger3}
Berger, B. K., Garfinkle, D., and Moncrief, V.  (1997), Numerical Study of Cosmological
Singularities,  gr-qc/9709073. 

\bibitem{berger4}
Berger, B. K., Garfinkle, D., and Strasser, E.  (1997), New
Algorithm for Mixmaster Dynamics, {\it Class. Quantum Grav.}, {\bf 14}, L29--L36. 

\bibitem{berger5}
Berger,
B. K. and Moncrief, V.  (1993), Numerical Investigation of Cosmological Singularities, {\it
Phys. Rev. D}, {\bf 48}: 4676. 

\bibitem{berger6}
Berger, B. K. and Moncrief, V.  (1997), Numerical Evidence
for Velocity Dominated Singularities in Polarized $U(1)$ Symmetric Cosmologies,
unpublished. 

\bibitem{gowdy}
Gowdy, R. H.  (1971), Gravitational Waves in Closed Universes, {\it Phys. Rev.
Lett.}, {\bf 27}:  826. 

\bibitem{grubisic1}
Grubi\u{s}i\'{c}, B. and Moncrief, V.  (1993), Asymptotic Behavior of the
$T^3  \times  R$ Gowdy Space-times, {\it Phys. Rev. D}, {\bf 47}, 2371--2382. 

\bibitem{grubisic2}
Grubi\u{s}i\'{c}, B.
and Moncrief, V.  (1994),  Mixmaster Spacetime, Geroch's Transformation, and Constants of
Motion, {\it Phys. Rev. D}, {\bf 49}, 2792--2800. 

\bibitem{hawking}
Hawking, S. W. and Penrose, R. (1970),
The Singularities of Gravitational Collapse and Cosmology, {\it Proc. Roy. Soc. Lond. A},
{\bf 314}:  529--548. 

\bibitem{isenberg}
Isenberg, J. A. and Moncrief, V. (1990), Asymptotic Behavior of the
Gravitational Field and the Nature of Singularities in  Gowdy Spacetimes,  {\it Ann. Phys.
(N.Y.)}, {\bf 199}: 84. 

\bibitem{kirillov}
Kirillov, A. A. and Kochnev, A. A.  (1987), Cellular Structure of
Space near a Singularity in Time in Einstein's Equations, {\it JETP Lett}., {\bf 46},
435--438. 

\bibitem{misner}
Misner, C. W. (1969), Mixmaster Universe, {\it Phys. Rev. Lett.}, {\bf 22}: 
1071--1074. 

\bibitem{moncrief1}
Moncrief, V.  (1981), Global Properties of Gowdy Spacetimes with $T^3 \times R$
Topology, {\it Ann. Phys. (N.Y.)}, {\bf 132}, 87--10. 

\bibitem{moncrief2}
Moncrief, V.  (1986), Reduction of
Einstein's Equations for Vacuum Space-Times with Spacelike $U(1)$ Isometry Groups, {\it
Ann. Phys. (N.Y.)}, {\bf 167}:  118. 

\bibitem{moser}
Moser, A.R., Matzner, R.A. and Ryan,  M.P., Jr. 
(1973), Numerical Solutions for Symmetric Bianchi Type IX Universes, {\it Ann. Phys.
(N.Y.)}, {\bf 79}, 558--579. 

\bibitem{rugh}
Rugh, S.E. and Jones, B.J.T.  (1990), Chaotic Behaviour and
Oscillating Three-Volumes in Bianchi IX Universes, {\it Phys. Lett.}, {\bf A147}, 353--359.

\bibitem{ryan}
Ryan,  M. P., Jr.  (1971), Qualitative Cosmology: Diagrammatic Solutions for Bianchi Type
IX Universes with Expansion, Rotation, and Shear.  I.  The Symmetric Case,  {\it Ann. Phys.
(N.Y.)}, {\bf 65}, 506--537.

\bibitem{weaver}
Weaver, M., Isenberg, J., and Berger, B.K.  (1997), Mixmaster Behavior in Inhomogeneous
Cosmological Spacetimes, gr-qc/9712055.

\end{references}
\end{document}